\documentclass[referee]{raa}
\usepackage{graphicx,times}             
\input{epsf.sty}                        
\input{psfig.sty}                       

\begin{document}
\title{The Phase Shifts of the Paired Wings of Butterfly Diagrams}
\volnopage{Vol.0 (200x) No.0, 000--000} \setcounter{page}{1}
\author{K. J. Li \inst{1,2}
\and H. F. Liang \inst{3}
\and W. Feng \inst{4}
   }
\institute{National Astronomical Observatories/Yunnan Observatory,
CAS, Kunming 650011, China
           {\it lkj@ynao.ac.cn}\\
           \and
Key Laboratory of Solar Activity, National Astronomical
Observatories, Chinese Academy of Sciences  \\
\and Department of Physics, Yunnan Normal University, Kunming 650093, China
\and  Research Center for Analysis and Measurement, Kunming University of
Science and Technology, Kunming 652094, China
}


\abstract{Sunspot groups observed by Royal Greenwich Observatory/US Air
Force/NOAA from  May 1874 to November 2008 and the Carte Synoptique
solar filaments from March 1919 to December 1989 are used to
investigate the relative phase shift of the paired wings of butterfly
diagrams of sunspot and filament activities. Latitudinal migration
of sunspot groups (or filaments) does asynchronously occur in
the northern and southern hemispheres, and there is a relative phase shift
between the paired wings of their butterfly diagrams in a cycle, making the
paired wings spatially asymmetrical on the solar equator.
It is inferred that hemispherical solar activity
strength should  evolve in a similar way within the paired
wings of a butterfly diagram in a cycle, making the paired wings
just and only keep the phase relationship between the northern and
southern hemispherical solar activity strengths, but a relative
phase shift between the paired wings of a butterfly diagram should
bring about an almost same relative phase shift of hemispheric solar
activity strength.
 }

\authorrunning{K. J. Li}
\titlerunning{Phase of the Paired Wings of Butterfly}
\maketitle
%
\section{Introduction}           
\label{sect:intro}
Sunspots are distributed themselves on the Sun with  a complex
spatial and temporal behavior. As for the spatial (latitudinal)
evolutional behavior, they display an evolution in the course
of a solar cycle to form a Maunder ``butterfly diagram"
(Maunder 1904, 1913, 1922; Hathaway 2010). As for the long-term temporal
evolutional behavior of sunspots' occurrence, the widely known
feature is their approximately 11-year Schwabe cycle (Schwabe
1843; Carrington 1858). It is found that the paired wings of a
Maunder ``butterfly diagram" are different from each other,
that is the well-known north\,--\,south asymmetry
of solar activity (Newton $\&$ Milson 1955; Li et al. 2002b;
Carbonell et al. 2007). Solar activity is found slightly
asynchronous in period phase between the solar northern and southern
hemispheres (Zolotova $\&$ Ponyavin 2006, 2007; Donner
$\&$ Thiel 2007; Li 2008), and the north\,--\, south
asymmetry of solar activity is related to the relative phase shifts
of solar activity in the northern and southern hemispheres
(Waldmeier 1957, 1971; Temmer et al. 2002, 2006). Phase
shifts (or phase differences) between the northern and southern
hemispherical solar activity strength should have a consequence: the
hemisphere preceding in time is more active at the ascending branch
of a sunspot cycle, whereas at the descending branch it is the
hemisphere following in time (Waldmeier 1971).
However, the
minima of sunspot activity are usually in phase, which might reveal
a kind of ``cross-talk" between the northern and southern
hemispheres at the end of a solar cycle (Temmer et al. 2006),
that is to say, phase shifts should have solar activity strength at the end
of one cycle in the preceding hemisphere to overlay that at the
beginning of the next cycle in the following hemisphere. Using the
monthly number of sunspot groups respectively in the northern and
southern hemispheres in cycles 12 to 23, the number of filaments
respectively in the two hemispheres in Carrington rotations 876 to
1823 covering cycles 16 to 21, the monthly mean northern and
southern hemispheric sunspot numbers in cycles 19 to 23, the monthly
mean northern and southern hemispheric sunspot areas in cycles 12 to
23, and the monthly mean northern and southern hemispheric flare
indices in cycles 20 to 23, Li (2009) found that solar
activity strength does asynchronously occur in the northern and southern
hemispheres, and there is a systematic time delay between the two
hemispheres in a cycle. It should be emphasized that the above five
solar indices reveal some kinds of solar activity ``strength"
(amplitude) varying with time.

A relative phase shift of hemispherical solar activity strength should have
an other consequence: ``the distance from the equator of the zone of
activity is smaller during the whole cycle for the hemisphere
preceding in time than for the hemisphere following in time"
(Waldmeier 1971). However, to validate such a consequence
it must be assumed that a relative
phase shift of hemispheric solar activity strength
should be kept in the paired wings of a butterfly diagram,
namely, a relative phase shift of the paired time series
(hemispherical solar activity strength) should exist
in the corresponding spacial distribution of the paired wings.

Solar activity strength is usually embedded into
butterfly diagrams of solar activity. A systematic time delay
between the northern and southern hemispheric solar activity
strengths (e.g. the aforementioned  five indices) in a cycle does
not indicate the existence of a relative shift in the paired wings
of the corresponding butterfly diagram of solar activity in the
cycle, a spacial (latitudinal) distribution of solar activity.
Even if the paired wings of a butterfly diagram of solar activity in a
cycle have no relative shift, a systematic time delay can exist
between the hemispheric solar activity strengths in the cycle, and
vice versa. That is to say, we can not infer whether a phase lag (or
lead) exist in a pair of wings of solar activity from a known phase
lag (or lead) of the corresponding hemispherical solar activity
strength, and vice versa.
Relative phase shifts between the paired time series of hemispheric
solar activity strength may be independent to relative phase shifts between
the paired wings of the corresponding butterfly diagram of solar activity, and
the above assumption should  not be spontaneously true.
Therefore, it is still needed to
investigate phase shifts in the paired wings of butterfly diagrams
of solar activity, although phase shifts of hemispherical solar
activity strength  have been already investigated. Thus, in the
present study we will directly investigate relative phase shifts in the
paired wings of butterfly diagrams of both sunspot and filament
activities, and further compare them with the relative phase shifts in the
hemispherical sunspot and filament activity strengths.

\section{The relative phase shifts of the paired wings of butterfly
diagrams}
\subsection{Sunspot butterfly diagram}
The observational data of sunspot groups used in the present study
come from Royal Greenwich Observatory/US Air Force/NOAA sunspot
record data
set\footnote{http://solarscience.msfc.nasa.gov/greenwch.shtml}.
The used data set comprises sunspot groups during the period of  May
1874 to November 2008 and covers solar cycles 12 to 23. Based on the
data set, a new data set is generated, in which each of all sunspot
groups is counted once (reserved is the first appearance of a
sunspot group in the new database), even though it was recorded
several times (or more) in the old data set, because it was observed
in several days (or more) when it passed through the solar disk. We
plotted individual sunspot groups in the new data set in the
latitude\,--\,time coordinate system. Figure 1 shows the resulting
latitudinal drift of sunspot group occurrence, which is called the
butterfly diagrams of sunspot groups. The figure obviously shows
some features, such as sunspot group occurrence in two zones
parallel to the solar equator whose latitudes are hardly greater
than $50^{\circ}$. The difference in appearance between the first
ten butterfly diagrams and the last three is due to the fact that
latitude values of sunspot groups are expressed in one decimal digit
before the year 1976, but since then they are given without any
decimal digit.

\vspace{10mm}
\begin{figure}
\noindent
\includegraphics[width=\textwidth, angle=0]{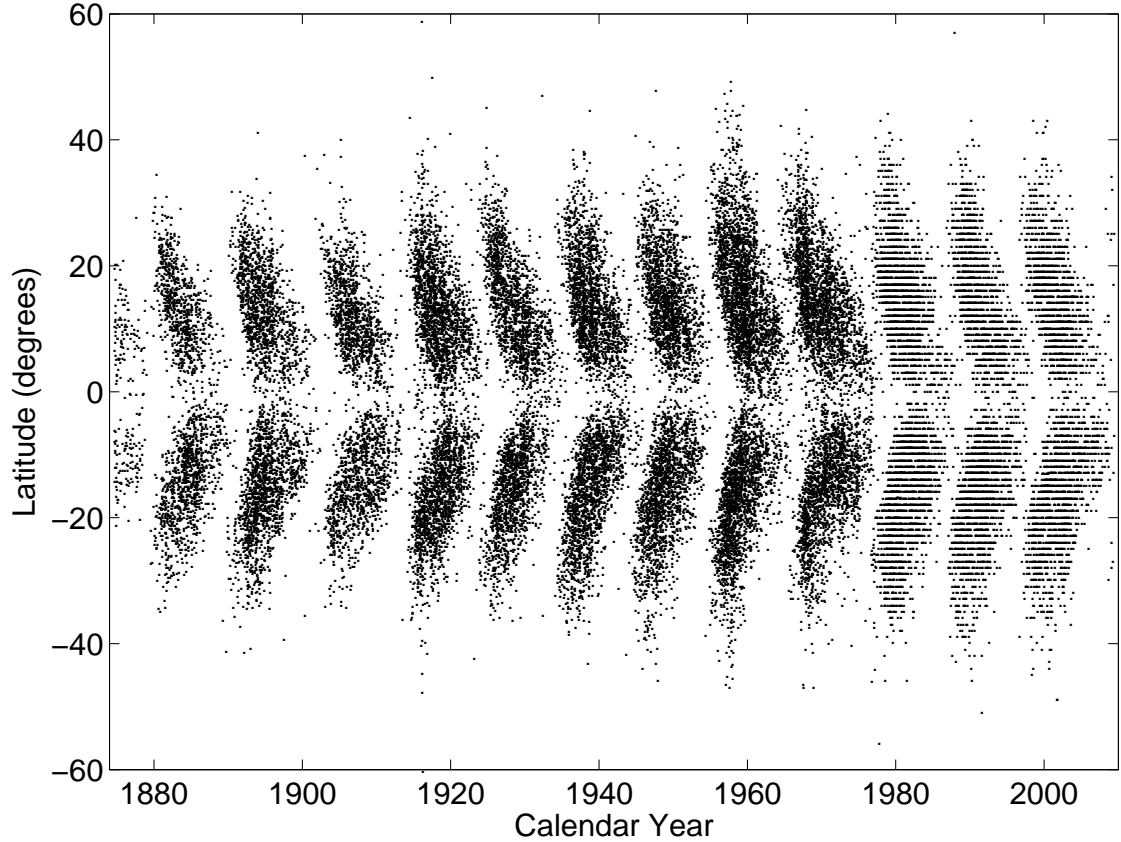}
 \caption{Butterfly diagrams of
sunspot groups during the period from May 1874 to November 2008,
coming from Royal Greenwich Observatory/US Air Force/NOAA sunspot
record data set.}
\end{figure}

It is difficult to accurately divide sunspot groups into solar
cycles to which they really belong (Harvey 1992). According
to the criterion for dividing sunspots into associated cycles,
proposed by Li et al. (2001) (in which
latitudes should be a function of time from sunspot cycle minimum),
sunspot groups
are roughly divided into individual butterflies. The monthly mean latitudes
of sunspot groups  respectively in the northern and southern
hemispheres,  marked respectively by {\bf $\stackrel{\rightarrow}{X_{n}}$} and
{\bf $\stackrel{\rightarrow}{X_{s}}$},
are calculated and then plotted in Figure 2.
Their corresponding standard errors are also calculated and
marked respectively by {\bf $\stackrel{\rightarrow}{\sigma_{n}}$} and
{\bf $\stackrel{\rightarrow}{\sigma_{s}}$}.
Although we distinguish the hemispheric labeling of the monthly mean
latitudes by different marks and colors in the
figure, a monthly mean
latitude at north $20^{\circ}$ or south $20^{\circ}$ is
plotted and will be used to calculate in the following as the same value of $20^{\circ}$ and so forth. The
criterion somewhat avoid the so-called ``cross-talk" of solar
activity between the northern and southern hemispheres at the end of
a cycle. Then we calculate
the average of  the absolute values of the differences
($AAD$) between the monthly mean latitudes of sunspot groups
respectively in the northern and southern hemispheres in each of
cycles 12 to 23. Next,
the wholly northern-hemispheric monthly mean latitudes in each cycle
are one-month shifted with
respect to the corresponding wholly southern-hemispheric monthly mean latitudes
along the calendar time axis, and then we get a new value of $AAD$.
Next again, the former are two-month shifted, and a new value of $AAD$ is obtained
again, and so on and so forth.
Resultantly, Figure 3 shows the average of  the absolute values of the differences
($AAD$) between the monthly mean latitudes of sunspot groups
respectively in the northern and southern hemispheres in each of
cycles 12 to 23, varying with relative phase shifts.
In the figure the abscissa indicates the relative
shift of the wholly northern-hemispheric monthly mean latitudes with
respect to the wholly southern-hemispheric monthly mean latitudes along the calendar time axis,
with negative values representing backward shifts.
When we do the above calculation, only those paired data are used, that is to say,
if only one hemisphere has one datum at a certain time, then the datum at the time is
not used to calculate the average.

Figure 4 shows the
relative shift corresponding to the minimum $AAD$ in each cycle.
In order to estimate error in the relative shift,
similarly, the time series {\bf $\stackrel{\rightarrow}{X_{n}} \pm \stackrel{\rightarrow}{\sigma_{n}}$}
vs {\bf $\stackrel{\rightarrow}{X_{s}} \pm \stackrel{\rightarrow}{\sigma_{s}}$}
are used to calculate their $AAD$, and then we get the
relative shift corresponding to the minimum $AAD$ in each cycle.
{\bf $\stackrel{\rightarrow}{X_{n}} \pm \stackrel{\rightarrow}{\sigma_{n}}$}
vs {\bf $\stackrel{\rightarrow}{X_{s}} \pm \stackrel{\rightarrow}{\sigma_{s}}$}
have 4 different pained combinations, finally giving 4
relative shifts in each cycle. Among the four relative shifts, the maximum (minimum) one corresponds
to the upper (lower) limit of an error bar, which is shown in Figure 4.
As the figure shows, the latitude migration of sunspot groups does not
synchronously occur in the northern and southern hemispheres, and
there is a relative shift between the paired wings of a butterfly
diagram in a cycle. Further, the relative shifts running from cycles
20 to 23 seem to repeat the shifts in cycles 12 to 15, implying a
possible period of about 8 cycles. In such a period, the relative
shifts dynamically drift from the obvious northern hemispheric lead
to the clear southern hemispheric lead. Also shown in the figure is
the systematic time delay between the monthly numbers of sunspot
groups respectively in the northern and southern hemispheres in each
of cycles 12 to 23 (Li 2009). As Figure 4 shows, the relative
shift between the monthly mean latitudes of sunspot groups
respectively in the northern and southern hemispheres in a cycle
seems to have a value very close to the systematic time delay
between the monthly numbers of sunspot groups  in the northern and
southern hemispheres in the cycle. It is thus inferred that
hemispherical solar activity strength should  evolve
in a similar way within the paired wings of a butterfly diagram in a cycle,
making the paired wings just and only embody (keep) the phase
relationship between the northern and southern hemispherical solar
activity strengths, and a phase difference between the paired wings
of a butterfly diagram, which is shown here by a relative shift
between the northern and southern hemispheric latitude migrations,
should bring about an almost same relative phase shift of
hemispheric solar activity strength.

\vspace{10mm}
\begin{figure}
\noindent
\includegraphics[width=\textwidth, angle=0]{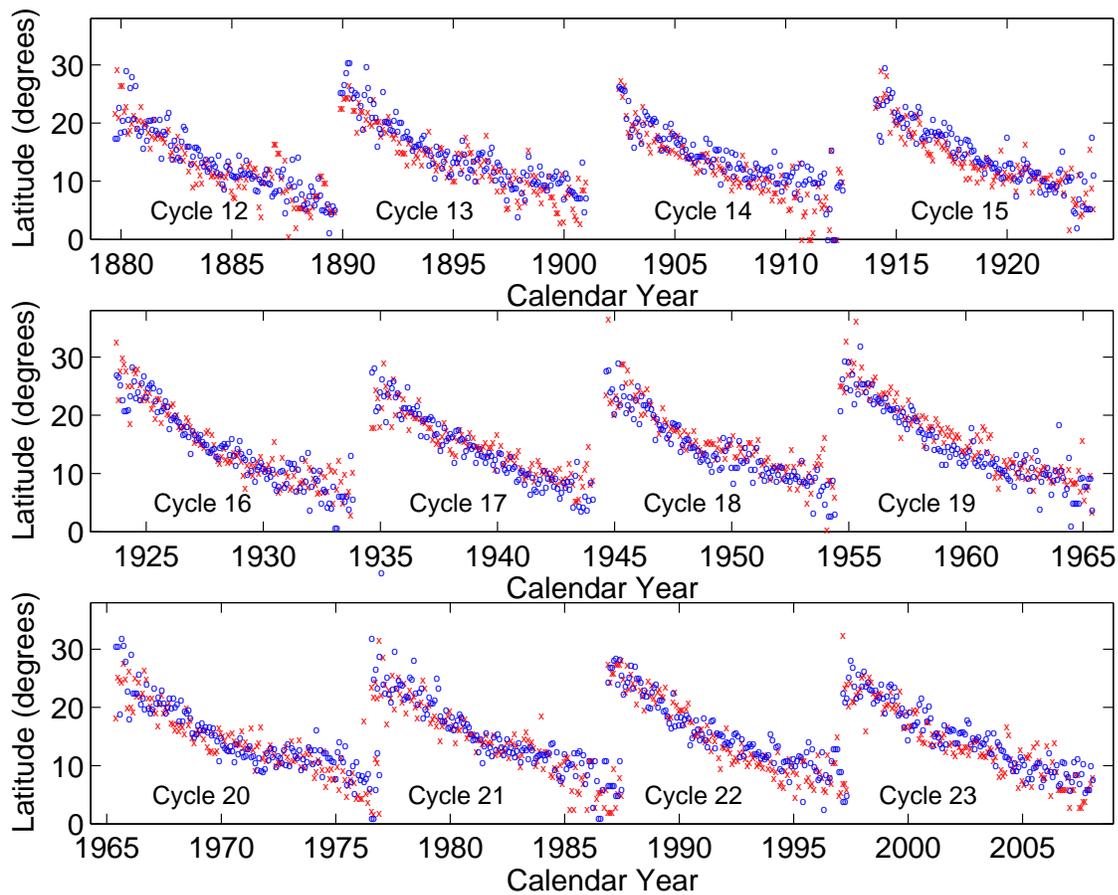}
 \caption{The monthly mean
latitudes of sunspot groups respectively in the northern (red
crosses) and southern (blue circles) hemispheres.}
\end{figure}

\vspace{10mm}
\begin{figure}
\noindent
\includegraphics[width=\textwidth, angle=0]{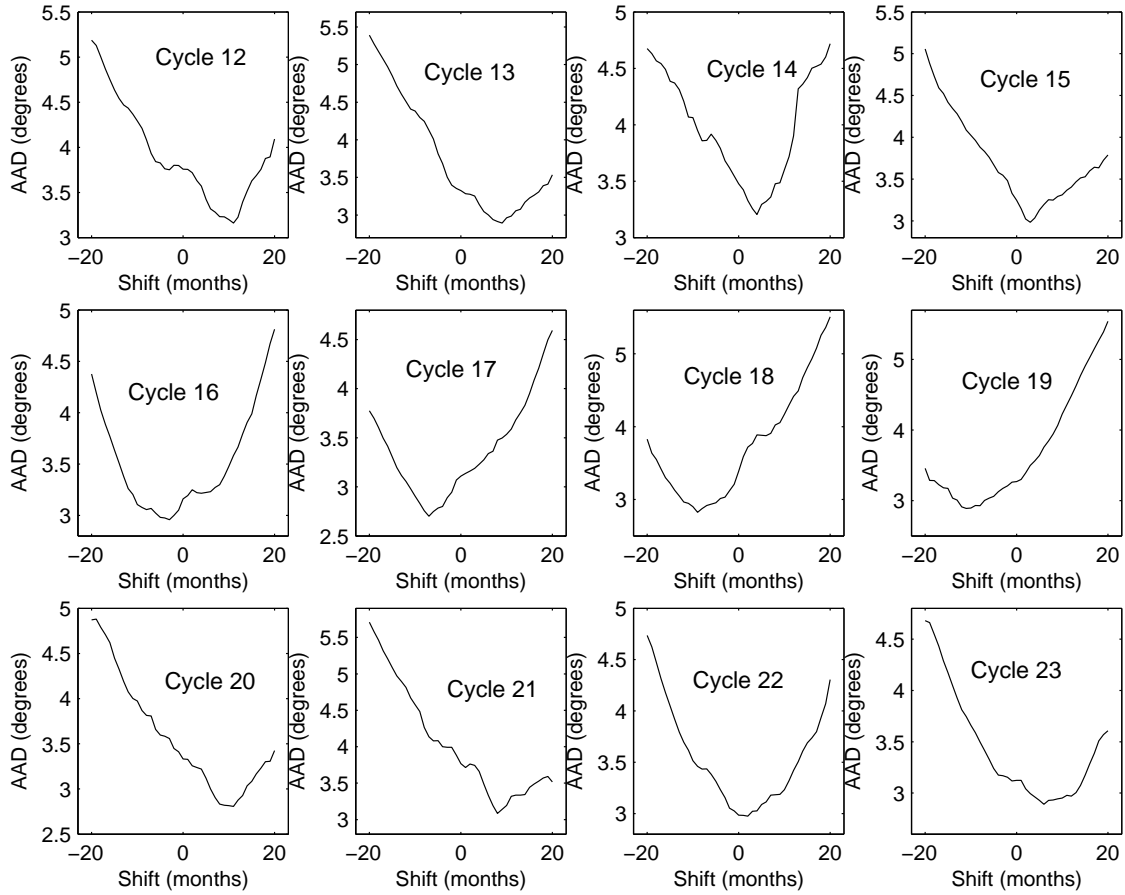}
 \caption{Average of the absolute values of the differences between the monthly
mean latitudes of sunspot groups in the southern and northern
hemispheres in each of cycles 12 to 23. The abscissa indicates the
shift of the wholly northern-hemispheric monthly mean latitudes in a cycle with
respect to the wholly southern-hemispheric monthly mean latitudes in the cycle
along the calendar time axis, with
negative values representing backward shifts.}
\end{figure}

\vspace{10mm}
\begin{figure}
\noindent
\includegraphics[width=\textwidth, angle=0]{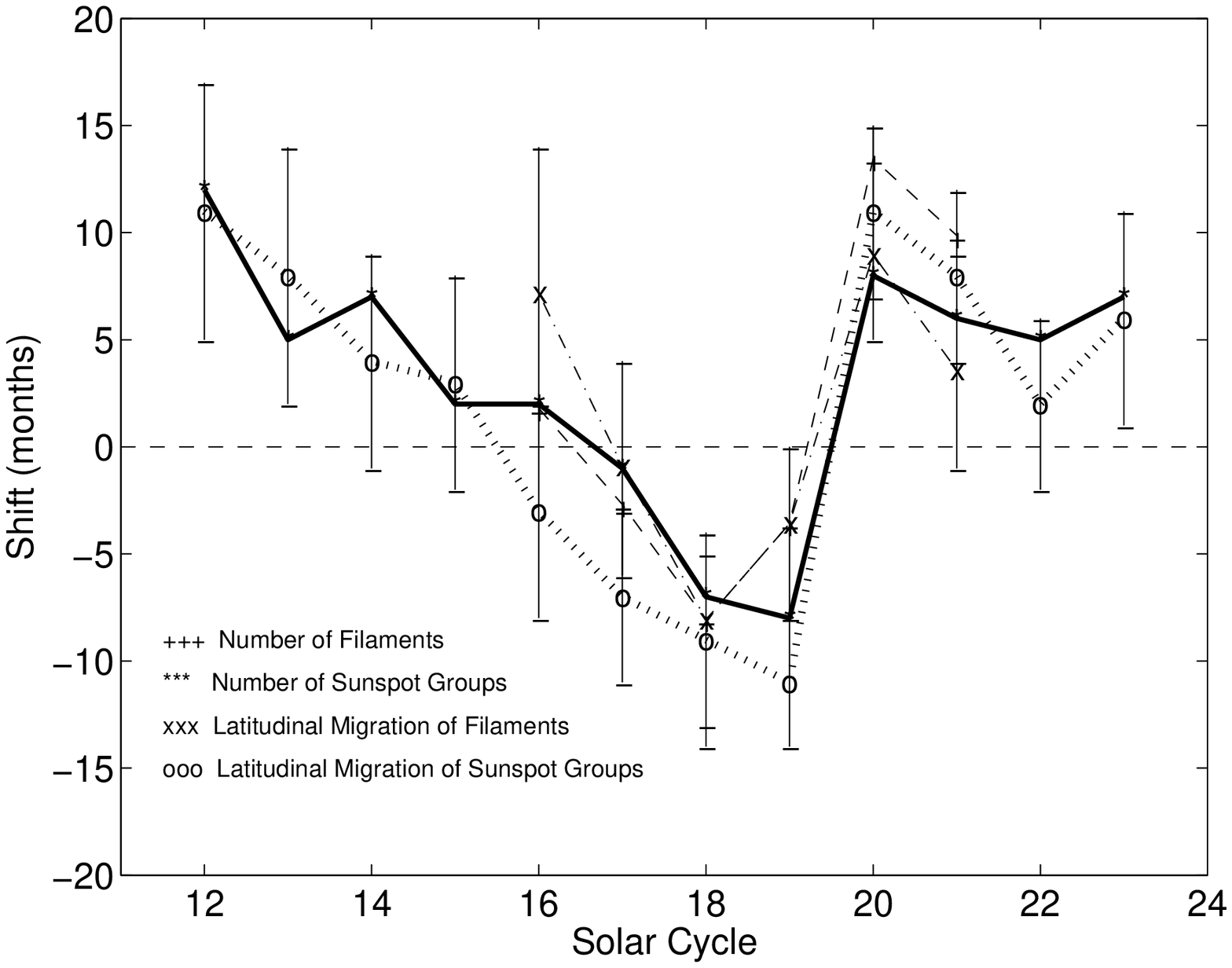}
 \caption{Relative shift corresponding to the minimum value of the
average of the absolute values of the differences ($AAD$) respectively of the
monthly mean latitude of sunspot numbers (circles)  and  the mean
latitude of filaments per Carrington rotation (crosses) in two solar hemispheres
in a cycle.
Their corresponding error bars are also displayed as thin solid vertical lines.
 Also shown
in this figure are the systematic time delay (asterisks) between the
monthly numbers of sunspot groups in the northern and southern
hemispheres in each of cycles 12 to 23 and that (plus signs) between
the  numbers of filaments per Carrington rotation  in the northern
and southern hemispheres in each of cycles 16 to 21 (Li
2009)}
\end{figure}

\subsection{Filament butterfly diagram}
Also utilized here is the Carte Synoptique solar filaments
archive\footnote{$ftp://ftp.ngdc.noaa.gov/STP/SOLAR_{-}DATA/SOLAR_{-}FILAMENTS$},
namely the catalogue of solar filaments from March 1919 to December
1989, corresponding to Carrington solar rotations 876 to 1823 and
covering 6 complete cycles from cycles 16 to 21 (Coffey $\&$
Hanchett 1998). The data of filaments span 948 Carrington
rotations, corresponding to 850 months, one Carrington solar
rotation is thus about 0.897 months. Using the data archive, we plot
the latitude drift of filament occurrence, which is called the
butterfly diagrams of filaments, shown in Figure 5. The normal solar
activity is usually applied to solar active events whose latitudes
are less than $50^{\circ}$ (Sakurai 1998; Li et al. 2002a).
Similarly, we count the
mean latitudes of filaments  whose latitudes are less than
$50^{\circ}$ in each of the considered Carrington rotations,
respectively in the northern and southern hemispheres. They are
shown in Figure 6.

\vspace{10mm}
\begin{figure}
\noindent
\includegraphics[width=\textwidth, angle=0]{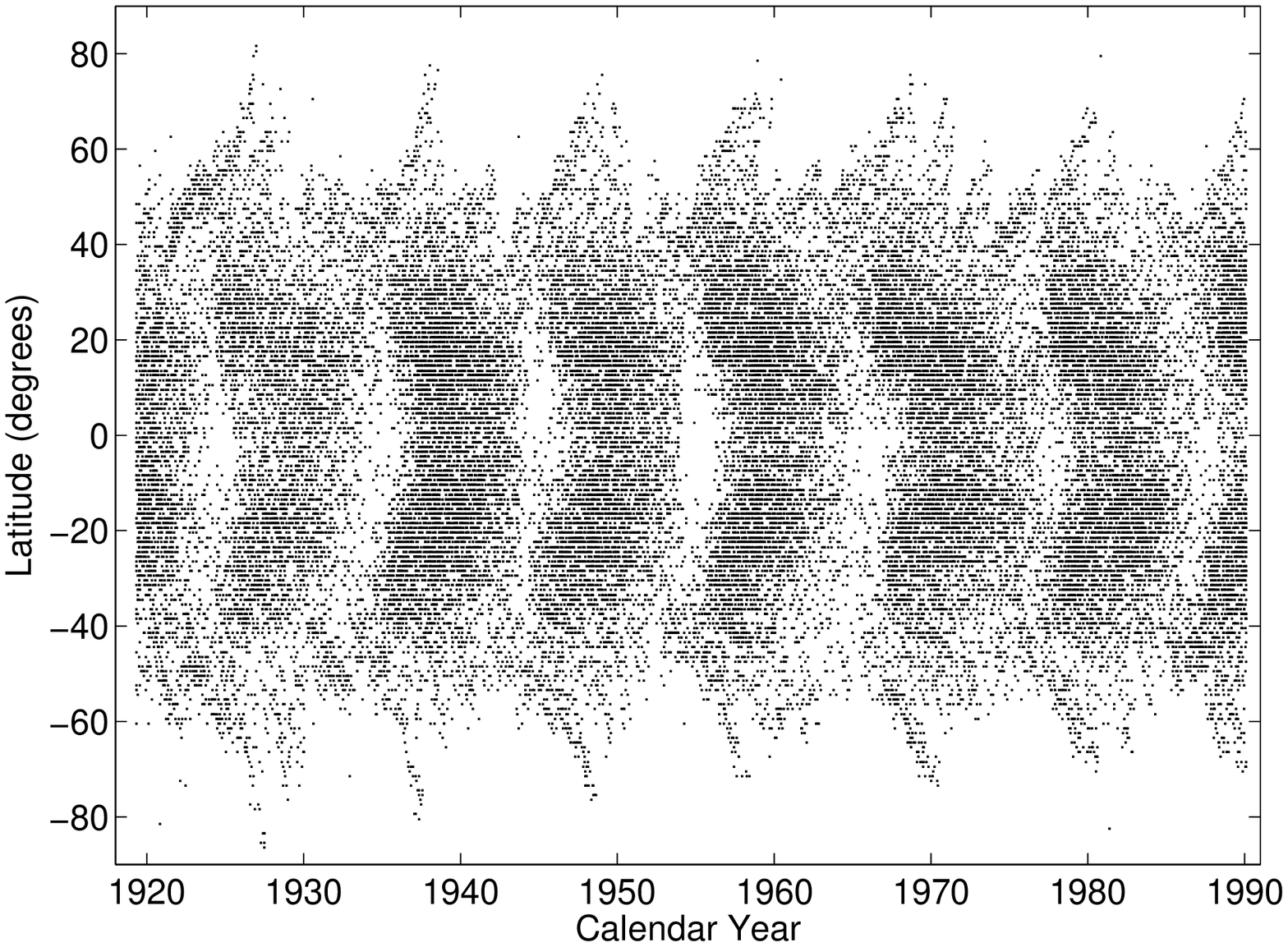}
 \caption{Butterfly diagram of
filaments from March 1919 to December
1989, namely from Carrington solar rotations 876 to 1823, coming from
the Carte Synoptique solar filaments archive.}
\end{figure}

\vspace{10mm}
\begin{figure}
\noindent
\includegraphics[width=\textwidth, angle=0]{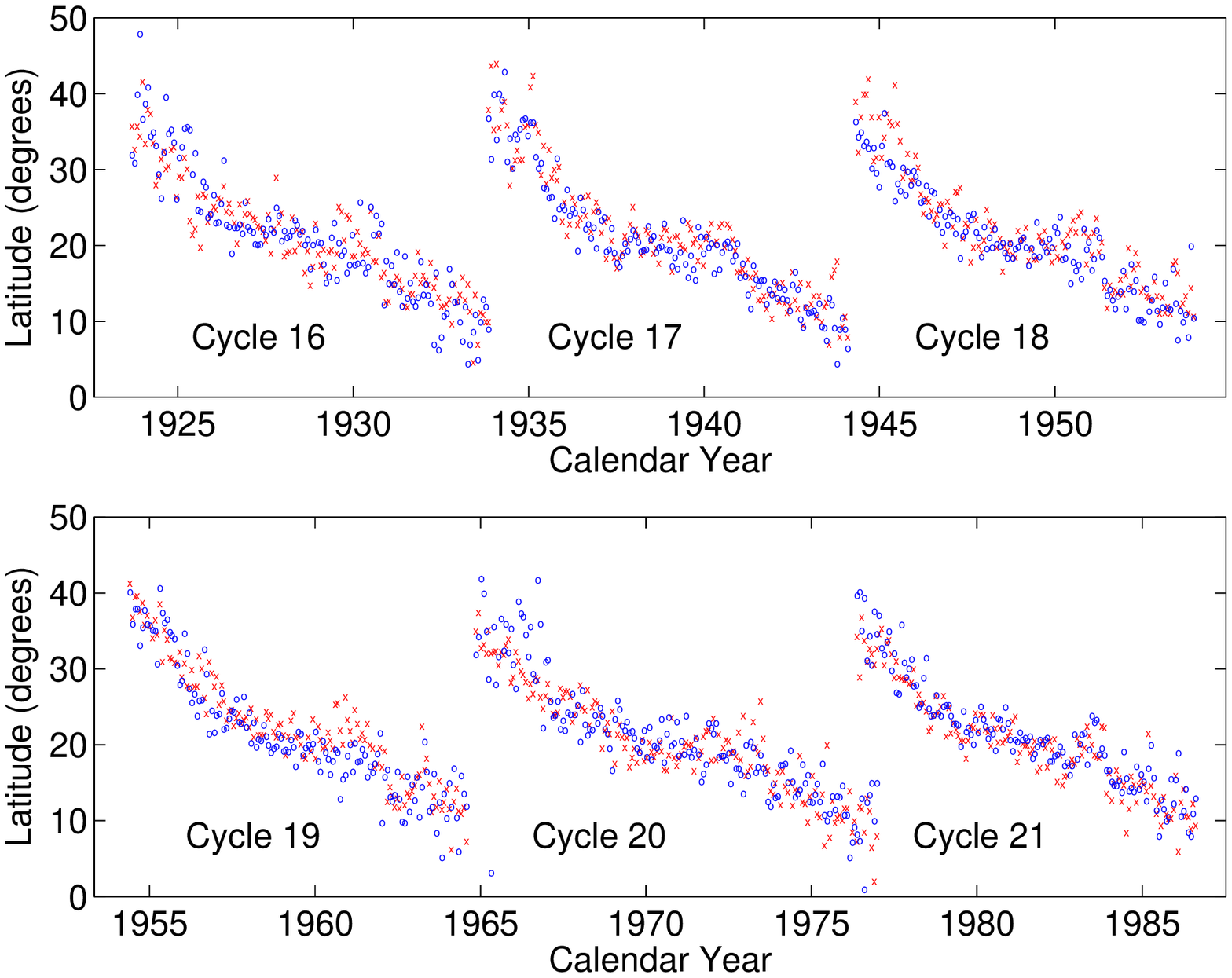}
 \caption{The mean latitudes
of filaments  whose latitudes are less than $50^{\circ}$ in each of
the considered Carrington solar rotations, respectively in the northern
(red crosses)   and  southern (blue circles) hemispheres. Carrington solar rotations
are translated into calendar times (in years).}
\end{figure}

Figure 7 shows the $AAD$ between the mean latitudes of filaments whose
latitudes are less than $50^{\circ}$  respectively in the  northern
and southern hemispheres in each of cycles 16 to 21. In the same way,
the abscissa in the figure indicates the shift of the northern-hemispheric  mean
latitudes with respect to the southern-hemispheric  mean latitudes,
with negative values representing backward shifts. The relative
shift corresponding to the minimum $AAD$ in a cycle is shown in Figure
4, and its error bar is also indicated in the figure, which is obtained
in the same way as for sunspot groups mentioned above.
As the figure shows, the latitude migration of filament activity
does not synchronously occur in the northern and southern
hemispheres, and there is a relative shift (systematic time lag or
lead) between the paired wings of a filaments' butterfly diagram in
a cycle. Also shown in the figure is the systematic time delay
between the numbers of filaments per Carrington rotation  in the
northern and southern hemispheres in each of cycles 16 to 21
(Li 2009). As Figure 4 shows, the relative shift between the mean
latitudes of filaments per Carrington rotation in the northern and
southern hemispheres in a cycle seems to have the almost same value
as the systematic time delay between the numbers of filaments per
Carrington rotation respectively in the northern and southern
hemispheres in the cycle. Thus, the inference given by the above
analysis to sunspot groups is valid for filament activity.

\vspace{10mm}
\begin{figure}
\noindent
\includegraphics[width=\textwidth, angle=0]{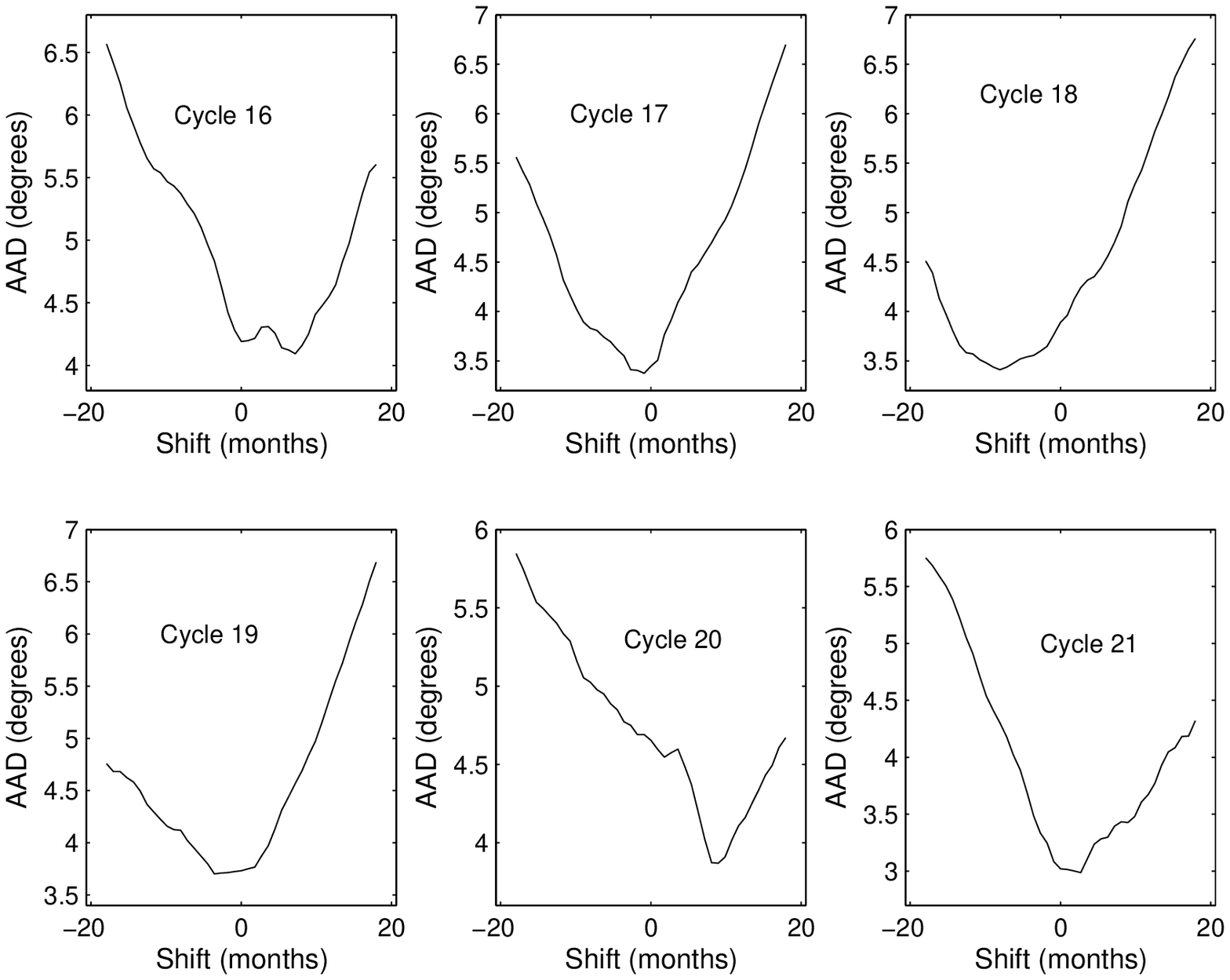}
 \caption{Average of the absolute values of the differences ($AAD$) between the mean
latitudes of filaments whose latitudes are less than $50^{\circ}$ in
the northern  and southern hemispheres  in each of cycles 16 to 21.
The abscissa indicates the shift of the wholly northern-hemispheric mean
latitudes in a cycle with respect to the wholly southern-hemispheric mean latitudes
in the cycle along the Carrington rotation time axis,
with negative values representing backward shifts.
Carrington solar rotations
are translated into calendar times (in years).
}
\end{figure}

\section{Conclusions and Discussions}
Using the data of sunspot groups observed by Royal Greenwich
Observatory/US Air Force/NOAA from  May 1874 to November 2008 and
the Carte Synoptique solar filaments  from March 1919 to December
1989, we have found that the latitudinal migration of hemispheric
solar activity (sunspot groups and filaments) does asynchronously
occur in the northern and southern hemispheres, and there is a
relative shift between the two hemispheres in a solar cycle, that is
to say, the paired wings of a butterfly diagram have a relative
shift between the northern and southern hemispheres along time
scale, making the paired wings spatially asymmetrical on the solar equator.
Further, for the latitudinal migration of both sunspot and
filament activities, phase shifts running from cycles 20 to 23 seem
to repeat those in cycles 12 to 15, implying the existence of a
possible period of about 8 cycles. Waldmeier (1971) once
analyzed the difference between the mean distance of the northern
sunspots to the solar Equator and that of the southern sunspots to
the Equator and also found the existence of a
possible period of about 8 cycles. The relative shift between
the monthly mean latitudes of sunspot groups (or filaments)
 in the northern and southern hemispheres in a cycle
seems to have an almost same value as the systematic time delay
between the monthly mean numbers of sunspot groups (or filaments) in
the northern and southern hemispheres in the cycle. It is thus
inferred that hemispherical solar activity strength of both sunspot
groups and filaments should evolve in a similar way  within
the paired wings of a butterfly diagram in a cycle, making the
paired wings just and only embody  the phase relationship between
the northern and southern hemispherical solar activity strengths,
and a phase difference between the paired wings of a butterfly
diagram, shown here by a relative shift between the northern and
southern hemispheric latitude migrations of sunspot groups or
filaments, should bring about an almost same relative shift of
hemispheric solar activity strength.
At the present, solar dynamo theory attempts to explain the north\,--\,south
asymmetry of solar activity strength (Goel $\&$ Choudhuri 2009), in which exists a
characteristic scale of about 12 cycles. In the future,
it is an important issue for solar dynamo theory to interpret
the relative phase shift of the paired wings of butterfly diagrams
the relative phase shift of the paired time series of hemispheric solar
activity strength.

Through wavelet scale-resolved phase coherence analysis of
hemispheric sunspot activity (the monthly mean numbers of sunspot
areas), Donner $\&$ Thiel (2007) gave in their Figure 4 the
phase difference between sunspot areas respectively in the northern
and southern hemispheres, which is the continuous phase shifts of
sunspot areas at frequency (period) band of 8 to 14 years. For
sunspot areas, phase differences are only coherent within a narrow
range of frequencies, which corresponds to time scales of about 8 to
14 years, therefore, phase coherence
is frequency-dependent. The continuous phase shifts are found
essentially similar to the relative shift of the hemispheric
latitude migration shown in Figure 4 and to that of the hemispheric
sunspot and/or filament activity strength (Li 2009), and  the
reason why the first is similar to the latter two is inferred to
that hemispheric solar activity periodically fluctuates with the
quasi 11-year cycle.  The phase difference between the paired wings
of a butterfly diagram of solar activity should lead to phase
asynchrony (shifts) of hemispheric solar activity strength, and it
should be an obvious reason which causes the asynchronization of
hemispheric solar activity strength.

Long-term observations of solar activity indicate that  solar
activity strength is asymmetrically distributed in the northern and southern
hemispheres, and the north \,--\, south difference (asymmetry) of
solar activity strength is a real phenomenon and not due to random
fluctuations (Li et al. 2009 and references therein), that is
to say, the paired wings of a butterfly diagram are different from
each other in activity strength.
The north-south asymmetry of sunspot latitudes has the same regularity as
that of sunspot numbers and areas (Pulkkinen et al.  1999; Li  et al.,
2002b), so, the hemisphere with more activity would have that activity at higher
latitudes.
A long-term characteristic time
scale of about 12 cycles should exist in the north \,--\, south
asymmetry of solar activity strength (and latitudes), and the dominant hemispheres of solar
activity strength (or hemispheres with higher latitudes) in a cycle regularly vary with solar cycles (Verma
1993; Li et al. 2002b), namely, long-term solar activity
strength and hemispheric relative average latitude regularly runs in the solar hemispheres with a possible
period of about 12 cycles. However the phase difference of the
paired wings (spacial distribution) of a butterfly diagram runs in the hemispheres with a
possible period of about 8 cycles. The cyclic variation of dominant
hemispheres of solar activity strength seems to have little relation with the
cyclic variation of the systematic time delay of solar activity
(strength or latitudinal migration),
and the systematic time delay of solar activity seems to
have little relation with hemispheric relative latitudes,
but within a cycle the
north-south asymmetry of solar activity strength may be strengthened
by  the systematic time delay between the northern and southern
hemispheric solar activity strengths (Waldmeier 1971;
Li 2009) or by the phase difference in the paired wings of the
butterfly diagram of solar activity in the cycle.

\begin{acknowledgements}
We thank the two referees for their careful reading of the
manuscript and constructive comments which improve the original
version of the manuscript.
The work is supported by the NSFC under Grants 10873032, 10921303, and
40636031,  the National Key Research Science Foundation
(2006CB806303), and the Chinese Academy of
Sciences.
\end{acknowledgements}

\label{lastpage}
\end{document}